\documentclass[11pt]{article}

\usepackage{amsmath}
\usepackage{amssymb}
\usepackage{natbib}
\usepackage{graphicx}

\title{Multivariate emulation of computer simulators:\\ model selection and diagnostics with application to a\\ humanitarian relief model}
\author{Antony Overstall\\
School of Mathematics \& Statistics,\\
University of Glasgow,\\
Glasgow,\\
UK\\
\\
David Woods\\ 
Statistical Sciences Research Institute,\\
University of Southampton,\\
Southampton,\\
UK\\
}
\date{}
\begin{document}
\maketitle
\renewcommand{\baselinestretch}{1}
\large

\parindent=0pt

\textbf{Abstract}\\
We present a common framework for Bayesian emulation methodologies for multivariate-output simulators, or computer models, that employ either parametric linear models or nonparametric Gaussian processes. Novel diagnostics suitable for multivariate covariance-separable emulators are developed and techniques to improve the adequacy of an emulator are discussed and implemented. A variety of emulators are compared for a humanitarian relief simulator, modelling aid missions to Sicily after a volcanic eruption and earthquake, and a sensitivity analysis is conducted to determine the sensitivity of the simulator output to changes in the input variables. The results from parametric and nonparametric emulators are compared in terms of prediction accuracy, uncertainty quantification and scientific interpretability.

\section{Introduction \label{INTRO}}

There are many systems in the physical, social and engineering sciences for which physical experimentation is infeasible or unaffordable. Some examples include investigations on ecosystems, infectious diseases, climate change, and galaxy formation (see \citealp{kaco2006}, for a number of case studies). In such situations, it is now common for the scientist or engineer to develop a \textit{simulator}, or computer model, that provides an approximation of the observed response from the physical system. In essence, the simulator is a deterministic or stochastic mathematical function that maps the inputs of a system to a prediction of its outputs. 

A simulator that has been successfully calibrated and validated, perhaps using physical data, can be employed for a number of tasks including prediction, optimisation, and sensitivity and uncertainty analyses \citep{Kennedy}. However, both calibrating and exploiting the simulator typically requires very many simulator evaluations. For complex problems, the computational expense of the simulator means brute-force approaches to these problems are infeasible, taking many hours, days or even weeks. Therefore, a fundamental step in understanding and using simulators is often the construction of a statistical \textit{emulator}, or meta-model, through a \textit{computer experiment} \citep{Sacks}. Here, the simulator is run at a carefully selected collection of combinations of the input variables and the resulting evaluations are treated as data to which a statistical model, the emulator, is fitted. The emulator can then be used to produce fast predictions of the output of the simulator for any values of the input variables, along with an associated measure of the prediction uncertainty. The emulator can then replace and supplement the simulator in both statistical calibration and scientific investigation. For more on computer experiments, see \citet{Santner}, \citet{fls}, and \citet{LevySteinberg2010}.

A Bayesian approach is very natural when constructing statistical emulators \citep{ohagan2006} with the chosen statistical model treated as a prior distribution on the simulator outputs and prediction, with associated uncertainty quantification, via the posterior predictive distribution (see Section~\ref{bayesframe}). Typically, a nonparametric Gaussian process (GP) regression model \citep{rasmussen} is employed; its advantages include flexibly adapting to the simulator evaluations and, for deterministic simulators, interpolating between data points. However, for some simulators, these advantages may be more than offset by the computational expense of estimating the GP model, and simpler and more computationally efficient models, such as multivariate linear regression, may be effective and more interpretable. Whatever statistical approach is taken to constructing the emulator, an important step is assessing its adequacy through formal statistical diagnostics \citep{Bastos}.

Frequently, each run of a simulator outputs a multivariate response, perhaps as a result of a time series or other dynamic process. The purpose of this paper is to present a Bayesian framework for covariance-separable emulation of multivariate simulators using parametric and nonparametric models and to develop novel model diagnostic procedures appropriate for such emulators. As part of our presentation, we unify the \textit{multivariate Gaussian process emulator} of  \citet{Conti} and the \textit{lightweight emulator} of \citet{Rougier}. Through an application to a simulator of a humanitarian relief mission, we demonstrate effective emulation, model selection and model checking for multivariate problems with a mixture of continuous and categorical input variables.  


\subsection{A humanitarian relief simulator with multivariate dynamic output}\label{SICILY}

Simulators have a long history of use in military and civilian emergency planning (see, for example, \citealp{ingber}). DIAMOND (DIplomatic And Military Operations in a Non-warfighting Domain; \citealp{TaylorLane}) is an emergency planning simulator for modelling peace support operations such as humanitarian relief and peace keeping. DIAMOND is mission-based, with high-level operational plans deconstructed into missions for individual units. It is able to model the actions and interactions between a wide range of agents, including military forces in non-warfighting roles, non-governmental organisations (NGOs), indigenous forces and civilians. A range of environmental and infrastructure features can also be varied. 

\begin{figure}
\centering
\includegraphics[scale=0.7, viewport = 77 305 518 587]{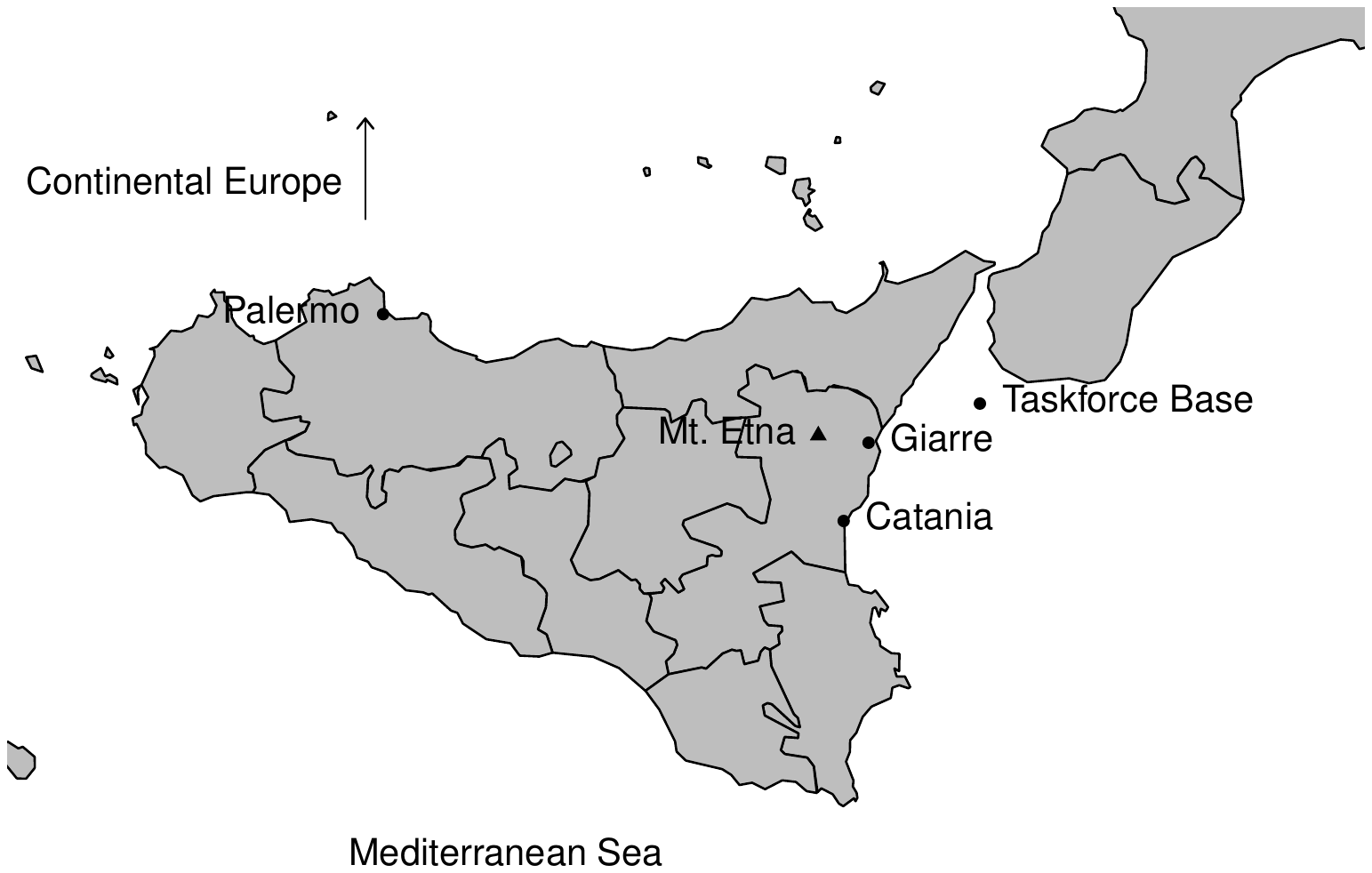}
\caption{\label{SIC_MAP}Map of Sicily showing the locations of Mount Etna, Giarre, Catania, a possible humanitarian task-force base, and the capital city Palermo}
\end{figure}

Our application of DIAMOND provides a deterministic model of a humanitarian relief mission to Sicily after an earthquake and subsequent eruption of Mount Etna. Etna is an active stratovolcano on the east coast of Sicily near the cities of Catania and Giarre (see Figure~\ref{SIC_MAP}). It has been designated a ``Decade Volcano'' by the International Association of Volcanology and Chemistry of the Earth's Interior and the United Nations due to its history of large eruptions and proximity to populated areas. Historically, more fatalities have been caused by earthquakes in the region, such as in 1693 when an earthquake of estimated magnitude 7.4 on the moment magnitude scale devastated the area and caused about 12,000 deaths in Catania ($\sim 63\%$ of the population at the time; \citealp{cfti}).   

The simulator models damage to the food supply, hospitals and housing (shelter) in Giarre and Catania resulting from the earthquake and eruption. An NGO launches a humanitarian relief operation which has two missions:

\begin{enumerate}
\item
\textit{Food Aid Mission}\\
To supply food to Catania and Giarre by helicopter from the NGO base.
\item
\textit{Repair Mission}\\
To transport engineers from the NGO base to Giarre and Catania, where they repair the food supply infrastructure and/or the shelter.
\end{enumerate}

\noindent We consider a scenario designed by the UK Defence Science and Technology Laboratory for the explicit and sole aim of model-testing; the scenario is not intended to support any real world decisions. Here, the NGO has four helicopter teams, two engineering teams and a single food depot. Two helicopter teams are assigned to the food aid mission and the others to transporting the engineers for the repair mission. 

The simulator has $p=13$ input variables, which represent the scale of the disaster and features of the humanitarian relief operation (see Table~\ref{SIC_INPUTS}). Eleven of these variables are continuous, with the other two being categorical with each having two levels. Input variables $x_1$-$x_6$ determine the impact of the earthquake and eruption on the population of Giarre and Catania by specifying the capacity of hospitals, shelter and food supply immediately following the disaster. The specification of these input variables creates a shortfall between population and shelter and/or food supply, leading to casualties. 

The remaining input variables (five continuous, two categorical) control certain features of the humanitarian relief mission. The continuous input variables are self-explanatory with the exception of $x_7$: the weighting of the engineer toolbox. This variable controls the relative importance given to repairing shelter and the food supply by the two engineering teams; $x_7=0$ ($x_7=1$) corresponds to engineers only repairing the shelter (food supply). 

\begin{table}
\caption{Input variables for the humanitarian relief mission simulator. The units of measurement for helicopter cargo capacity are specific to this simulator. Note that the initial populations in the simulator of Giarre and Catania are 27,000 and 300,000, respectively. Under normal circumstances, the simulator only expects 1\% of the population per day to require hospital treatment.\label{SIC_INPUTS}}
\centering
\begin{tabular}{llll} \hline
\multicolumn{4}{l}{Continuous input variables} \\ 
Name & Symbol & Range & Units \\ \hline
Giarre hospital capacity & $x_1$ & (135, 270) & person/day\\
Giarre shelter capacity & $x_2$ & (13500, 27000) & person/day\\
Giarre food supply capacity & $x_3$ & (13500, 27000) & person/day\\
Catania hospital capacity & $x_4$ & (2000, 3000) & person/day\\
Catania shelter capacity & $x_5$ & (200000, 300000) & person/day\\
Catania food supply capacity & $x_6$ & (200000, 300000) & person/day\\
Weighting of the engineer toolbox & $x_7$ & (0, 1) & N/A \\
Planning time for the humanitarian mission & $x_8$ & (36, 60) & hrs \\
Helicopter cruise speed & $x_9$ & (220, 270) & km/hr \\
Helicopter cargo capacity & $x_{10}$ & (7000, 7500) & N/A \\
Engineer ground speed & $x_{11}$ & (0, 10) & km/hr \\ \hline
\multicolumn{4}{l}{Categorical input variables} \\
Name & Symbol & \multicolumn{2}{l}{Levels} \\ \hline
Recipient of food aid & $x_{12}$ & \multicolumn{2}{l}{\{Giarre \& Catania, Catania only\}} \\
Location of NGO base & $x_{13}$ & \multicolumn{2}{l}{\{Continental Europe, Task-force Base\}} \\ \hline
\end{tabular}
\end{table}

The two levels for categorical variable $x_{12}$ correspond to, respectively, supplying food aid to both Giarre and Catania or to Catania alone. Although the second option is perhaps morally and politically unappealing, it may be practically relevant as there can be a much greater shortfall between the available and required food in Catania. Simulation modelling allows investigation of the impact of potentially unattractive options. For $x_{13}$, the two levels correspond to the NGO base being (i) in continental Europe or (ii) part of a military task force located on a fleet of ships in in the Strait of Messina between Italy and Sicily (see Figure~\ref{SIC_MAP}).

Each run of the simulator is defined by a setting for $x_1$-$x_{13}$. The output from each simulator run is the number of civilian casualties that have occurred on each of days 2,3,4,5 and 6 following the disaster. Therefore, the output for each run is a five dimensional vector.

\subsection{Bayesian emulators}
\label{bayesframe}

A Bayesian approach will be taken to constructing an emulator for the DIAMOND simulator. Let $\mathbf{x}=\left(x_1,...,x_p\right)^\mathrm{T} \in \mathcal{X} \subset \mathbb{R}^p$ denote the vector of $p$ input variables, with $\mathcal{X}$ the $p$-dimensional input space. The simulator is assumed to be a black-box function, $f:\mathcal{X} \to \mathcal{Y} \subset \mathbb{R}^k$, with $\mathcal{Y}$ the $k$-dimensional output space; that is

$$f(\mathbf{x}) = \left\{f_1(\mathbf{x}),...,f_k(\mathbf{x})\right\}^\mathrm{T}\,,$$

\noindent is the $k \times 1$ output vector from the simulator at input combination $\mathbf{x}$. An emulator for $f(\cdot)$ is a prediction equation that provides a surrogate for $f(\mathbf{x}_0)$, where $\mathbf{x}_0$ is an input combination at which the simulator has not previously been evaluated.

For a collection of input combinations $\zeta=\left\{\mathbf{x}_1,...,\mathbf{x}_n\right\}$, with $\mathbf{x}_i = \left(x_{i1},...,x_{ip}\right)^\mathrm{T}$, the simulator outputs are collated into an $n\times k$ output matrix

$$Y = \left(
\begin{array}{c}
f(\mathbf{x}_1)^\mathrm{T}\\
\vdots \\
f(\mathbf{x}_n)^\mathrm{T}
\end{array}
\right)\,.$$

\noindent A priori, we assume that $Y$ is a realisation from a probability distribution, specified up to a $d \times 1$ vector of unknown parameters $\boldsymbol{\theta} \in \Theta$, with $\Theta\subset\mathbb{R}^d$ the parameter space. After running the simulator for the input combinations in $\zeta$, the emulator is constructed as the posterior predictive distribution (see, for example, \citealt{OHagan}, p. 89) of $\mathbf{y}_0=f(\mathbf{x}_0)$, given by

\begin{equation}
\pi(\mathbf{y}_0|Y) = \int_{\Theta} \pi(\mathbf{y}_0|\boldsymbol{\theta},Y) \pi(\boldsymbol{\theta}|Y)\, \mathrm{d}\boldsymbol{\theta}\,.
\label{POSTPREDDIST}
\end{equation}

\noindent Here, $\pi(\boldsymbol{\theta}|Y)$ is the posterior density function for $\boldsymbol{\theta}$, found using Bayes theorem, and $\pi(\mathbf{y}_0|\boldsymbol{\theta},Y)$ is the conditional posterior predictive density for $\mathbf{y}_0$.

In the remainder of this article,  methodology for multivariate Bayesian emulation is developed and applied. In Section~\ref{METHODOLOGY}, the detailed methodology used to obtain the posterior predictive distribution is described for both multivariate Gaussian processes and linear models. In Section~\ref{DIAGS}, model selection and diagnostics for multivariate emulators are developed and discussed. In Section~\ref{RESULTS}, results are presented from applying the methodology to emulating the DIAMOND simulator. Section~\ref{DISC} gives a brief discussion.

Code to fit the emulators described in this paper and the training and test datasets are provided as supplementary material. 

\section{Multivariate emulation via the posterior predictive distribution} 
\label{METHODOLOGY}

In this section, the posterior predictive distribution is derived for a general class of multivariate linear models that includes Gaussian process (GP) models and linear regression models. As such, the multivariate GP emulator of \citet{Conti} and lightweight emulator of \citet{Rougier} are special cases. We also demonstrate how the multivariate GP emulator can include categorical input variables using the distance metrics of \citet{Qian1}.   


Our basic modelling assumption is that any finite set of multivariate responses has a joint matrix normal distribution \citep{Dawid} with mean function a linear combination of unknown model parameters and a separable covariance structure with, potentially, correlations between outputs from the same run and also between different runs of the simulator. That is, for $n\times k$ response matrix $Y$

\begin{equation}
Y| B,\Sigma, A \sim \mathrm{MN}_{n,k} \left(HB,\Sigma,A\right)\,,
\label{DISTRIBUTION}
\end{equation}

\noindent where $HB$ is the $n\times k$ mean matrix and $\Sigma$ and $A$ are, respectively, $k\times k$ and $n\times n$ positive definite column and row scale matrices. Note that

$$\mathrm{vec}(Y)|B,\Sigma,A \sim \mathrm{N}_{nk} \left( \mathrm{vec}(HB),\Sigma \otimes A \right)$$

\noindent is a multivariate normal distribution, where $\mathrm{vec}(\cdot)$ denotes the vectorisation function that stacks columns of a matrix and $\otimes$ denotes the Kronecker product.

In~(\ref{DISTRIBUTION}), the matrix $H$ is the $n \times m$ model matrix with $i$th row given by $h(\mathbf{x}_i)^T$, where $h: \mathcal{X} \to \mathcal{H} \subset \mathbb{R}^m$ is a known function of the simulator inputs ($i=1,\hdots,n$). For example, if $h(\mathbf{x}) = (1,x_1)$, then the model contains an intercept and a linear term in $x_1$. If some input variables are categorical, then we define the appropriate elements of $h(\mathbf{x}_i)$ through the usual constraints, for example corner-point or sum-to-zero. The matrix $B$ is an $m \times k$ matrix of unknown regression parameters. 

The separability of the covariance structure implied by this matrix normal distribution results in a common scale matrix $\Sigma$ for the $k$ multivariate responses at each of the $n$ simulator runs. An emulator with a separable covariance structure is both easier to implement and interpret. If diagnostic measures (see Section~\ref{DIAGSUB}) suggest inadequacy of the separable emulator, alternative methodologies could be employed \citep[see, for example,][and references therein]{Fricker}.  


If homogeneity of variance across the simulator runs is assumed, that is $\mbox{Var}\left\{f(\mathbf{x}_i)\right\}=\Sigma$ for all $i=1,\ldots,n$, then $A$ can be specified as a correlation matrix. For the multivariate GP emulator, we define $A$ through a stationary correlation function, and set $ij$th entry equal to $a_{ij}=c(|\mathbf{x}_i-\mathbf{x}_j|;\mathbf{r})$, i.e. the correlation between any two rows of $Y$ depends only on the distance between $\mathbf{x}_i$ and $\mathbf{x}_j$ ($i,j=1,\ldots,n$) and a vector of unknown correlation parameters $\mathbf{r}$. The lightweight emulator is defined as a special case with 

$$
c(\mathbf{x}_i,\mathbf{x}_j;\mathbf{r}) = \left\{ \begin{array}{cl}
1 & \mbox{if $i=j$}\,,\\
0 & \mbox{if otherwise}\,. \end{array} \right.
$$

\noindent Thus we can replace conditioning on $A$ in~\eqref{DISTRIBUTION} by conditioning on $\mathbf{r}$. 

We use the conditionally conjugate (given $\mathbf{r}$) matrix-normal-inverse-Wishart (MNIW) prior distribution for $B$ and $\Sigma$, denoted 
$\mathrm{MNIW}_{m,k}\left(M,\Omega,S,\delta\right)$, where

\begin{eqnarray}
B|\Sigma,\mathbf{r} & \sim & \mathrm{MN}_{m,k}\left(M,\Sigma,\Omega\right),\label{SICILY_PRIOR1}\\
\Sigma|\mathbf{r} & \sim & \mathrm{IW}_k \left(S,\delta \right)\,. \label{SICILY_PRIOR2}
\end{eqnarray}
Here, $\mathrm{IW}_k$ denotes the inverse-Wishart distribution for $k \times k$ positive-definite matrices, $M$, $\Omega$ and $S$ are the $m\times k$, $m\times m$ and $k \times k$ matrices of hyperparameters, respectively, and $\delta>0$ is the prior degrees of freedom. The corresponding probability density function is given in Section~1 of the Supplementary Material, up to a normalising constant; see also \citet{Rougier}. 

Using this prior distribution the conditional posterior distribution, given $\mathbf{r}$,  is 

$$B,\Sigma |Y,\mathbf{r} \sim \mathrm{MNIW}_{m,k} \left(\hat{M},\hat{\Omega},\hat{S},\hat{\delta}\right)\,,$$

\noindent see Section~2 of the Supplementary Material, where

\begin{eqnarray*}
\hat{\Omega} & = & \left(H^\mathrm{T} A^{-1}H + \Omega^{-1}\right)^{-1}\,,\\
\hat{M} & = & \hat{\Omega} \left( H^\mathrm{T} A^{-1} Y + \Omega^{-1}M\right)\,,\\
\hat{S} & = & Y^\mathrm{T} A^{-1} Y + M^\mathrm{T} \Omega^{-1} M + S -  \hat{M}^\mathrm{T} \hat{\Omega}^{-1} \hat{M}\,,\\
\hat{\delta} & = & \delta + n\,.
\end{eqnarray*}

To predict the simulator output $Y_0 = \left[f(\mathbf{x}_{01}),\ldots,f(\mathbf{x}_{0n_0})\right]^\mathrm{T}$ at a set of $n_0$ test inputs, $\zeta_0 = \left\{\mathbf{x}_{01},...,\mathbf{x}_{0n_0} \right\}$, we first define the joint conditional distribution of $Y$ and $Y_0$

\begin{equation}
\left. \left( \begin{array}{c}
Y\\
Y_0 \end{array}\right) \right\vert B,\Sigma,\mathbf{r} \sim \mathrm{MN}_{n+n_0,k} \left( \left[ \begin{array}{c}
H\\
H_0 \end{array}\right]B, \Sigma,
\left[ \begin{array}{cc}
A & T \\
T^\mathrm{T} & A_0 \end{array}\right]\right)\,,
\label{DISTRIBUTION2}
\end{equation}

\noindent where $H_0$ is the $n_0 \times m$ matrix with $u$th row $h(\mathbf{x}_{0u})^T$, $A_0$ is the $n_0 \times n_0$ matrix with $uv$th element given by $c(\mathbf{x}_{0u},\mathbf{x}_{0v};\mathbf{r})$, and $T$ is the $n \times n_0$ matrix with $iu$th element given by $c(\mathbf{x}_i,\mathbf{x}_{0u};\mathbf{r})$ ($u,v=1,\ldots,n_0;\,i=1,\ldots,n$). 

It can be shown (see Section~3 of the Supplementary Material) that the conditional distribution of $Y_0$ is 

\begin{equation}
Y_0| Y, B, \Sigma,\mathbf{r} \sim \mathrm{MN}_{n_0,k} \left( H_0 B + T^\mathrm{T} A^{-1} (Y - HB), \Sigma, A_0 - T^\mathrm{T} A^{-1} T \right)\,.
\label{DISTRIBUTION3}
\end{equation}

From~\eqref{DISTRIBUTION2} and~\eqref{DISTRIBUTION3}, we can see the fundamental difference between the GP and lightweight emulators; for the lightweight emulator, the output from different simulator runs is assumed independent given $\left\{B,\Sigma\right\}$ and hence the matrix, $T$, of correlations between the observed and unobserved simulator runs will be a zero matrix. Hence, conditional on $B$ and $\Sigma$, the distribution of $Y_0$ does not depend on $Y$. For the GP emulator, with non-zero correlations between simulator runs, the dependence between $Y_0$ and $Y$ remains even after conditioning on $B$ and $\Sigma$.

To obtain the posterior predictive distribution of $Y_0$, given $\mathbf{r}$, we integrate~(\ref{DISTRIBUTION3}) with respect to the posterior distribution of $B$ and $\Sigma$ (see Section~4 of the Supplementary Material): 

\begin{equation}
Y_0 | Y, \mathbf{r} \sim \mathrm{MT}_{n_0,k} \left(Q, \hat{S},R,\hat{\delta}\right)\,,
\label{DISTRIBUTION4}
\end{equation}

\noindent where

\begin{eqnarray*}
Q & = & H_0 \hat{M} + T^\mathrm{T} A^{-1} \left(Y - H \hat{M}\right)\,,\\
R & = & A_0 - T^\mathrm{T} A^{-1} T + \left(H_0 - T^\mathrm{T} A^{-1} H\right) \hat{\Omega} \left(H_0 - T^\mathrm{T} A^{-1} H\right)^\mathrm{T}\,,
\end{eqnarray*}

\noindent and $\mathrm{MT}_{n_0,k}(Q,\hat{S},R,\hat{\delta})$ denotes the matrix t-distribution \citep{Javier} with location matrix $Q$, column scale matrix $\hat{S}$, row scale matrix $R$ and degrees of freedom $\hat{\delta}$. Marginal posterior predictive distributions for the $u$th simulator run, $\mathbf{y}_{0u}=f(\mathbf{x}_{0u})$, and the $s$th output, $y_{0,us}=f_s(\mathbf{x}_{0u})$ are multivariate and univariate $t$ distributions, respectively:

\begin{equation*}
\mathbf{y}_{0u}| Y, \mathbf{r} \sim \mathrm{t}_k \left(\mathbf{q}_u^T, \frac{R_{uu} \hat{S}}{\hat{\delta}}, \hat{\delta} \right)\,;
\label{DISTRIBUTION5}
\end{equation*} 

\begin{equation}
y_{0,us}| Y, \mathbf{r} \sim \mathrm{t}\left(q_{us},\frac{R_{uu}\hat{S}_{ss}}{\hat{\delta}},\hat{\delta}\right)\,.
\label{DISTRIBUTION6}
\end{equation} 

\noindent Here, $\mathbf{q}_u$ is the $u$th row of $Q$ and $q_{us}$ is the $us$th element of $Q$, $R_{uu}$ is the $u$th diagonal element of $R$ and $\hat{S}_{ss}$ is the $s$th diagonal element of $\hat{S}$. 

For the lightweight emulator, where $A=I_n$, an $n\times n$ identity matrix,~(\ref{DISTRIBUTION4}) provides closed-form posterior predictive distributions. For the multivariate GP emulator, and the most commonly used correlation functions $c(\cdot,\cdot;\mathbf{r})$, there does not exist a prior distribution for $\mathbf{r}$ such that a closed-form expression can be obtained for the marginal posterior predictive distribution of $Y_0$. Typically, one of two approaches is taken: (i) $\mathbf{r}$ is replaced by a ``plug-in'' estimate $\hat{\mathbf{r}}$, a representative value with respect to the marginal posterior distribution of $\mathbf{r}$; or (ii) Markov Chain Monte Carlo (MCMC) methods are used to sample from the marginal posterior distribution of $\mathbf{r}$ and then for each sampled value of $\mathbf{r}$, a value is drawn from the conditional posterior predictive distribution~(\ref{DISTRIBUTION4}).

The plug-in approach is less computationally expensive than the fully Bayesian approach and provides a closed-form emulator. We adopt the plug-in approach for prediction using the marginal posterior mode of $\mathbf{r}$, obtained by maximising the unnormalised marginal posterior density

$$\pi(\mathbf{r}| Y) \propto \pi_{\mathbf{r}}(\mathbf{r}) |A|^{-\frac{k}{2}}|\hat{\Omega}|^{\frac{k}{2}}|\hat{S}|^{-\frac{\hat{\delta}+k-1}{2}}\,,$$

\noindent where $\pi_{\mathbf{r}}(\mathbf{r})$ is the prior probability density function for $\mathbf{r}$. 

The final step in building the multivariate GP emulator is choice of the correlation function $c(\cdot,\cdot;\mathbf{r})$. The most commonly used function is the power exponential function, which was extended by \citet{Qian1} to incorporate both quantitative and qualitative variables. Assuming without loss of generality that the variables are ordered, so that the first $p_1$ variables in $\mathbf{x}$ are quantitative and the next $p-p_1$ are qualitative variables, a correlation function that is exchangeable in the levels of the qualitative variables has the form 

\begin{equation}
c(\mathbf{x}_1,\mathbf{x}_2;\mathbf{r}) = \exp \left\{ - \sum_{l=1}^{p_1} r_l |x_{1l} - x_{2l}|^{g_l} -\sum_{l=p_1+1}^p r_l \mathrm{I}(x_{1l} \ne x_{2l})\right\}\,. 
\label{CORR_2}
\end{equation}

\noindent \citet{Qian1} suggested a number of correlation functions for qualitative variables, each reducing to the common form~\eqref{CORR_2} for two-level qualitative variables. Throughout this paper, we fix $g_l=2$ for all $l$.

\section{Emulator diagnostics and improvement}
\label{DIAGS}

In this section, we address diagnostics for emulator adequacy and methods for improving emulator performance, including variable selection and the addition of a nugget term for the multivariate Gaussian process.

\subsection{Emulator diagnostics}
\label{DIAGSUB}

We start by developing generalisations to multivariate emulators of the diagnostics provided by \citet{Bastos} for univariate Gaussian process emulators. These diagnostics assess the assumption underlying~\eqref{DISTRIBUTION}, that the responses conditionally follow a matrix normal distribution with specified mean and correlation functions. Their evaluation requires an additional validation set of simulator runs, $\zeta_0$ and $Y_0$, to be available.

\subsubsection{Individual prediction errors}
\label{INDI}

As suggested by \citet{Bastos}, standardised prediction errors can be explored graphically or used to construct nominal-level predictive probability intervals. If the emulator is an adequate model of the simulator, from~(\ref{DISTRIBUTION6}), the standardised individual prediction error 

$$D_{us}^I(Y_0) = \sqrt{\frac{\hat{\delta}}{R_{uu}\hat{S}_{ss}}} \left(y_{0,us} - q_{us}\right)$$

\noindent has a standard t-distribution, conditional on $Y$ with $\hat{\delta}$ degrees of freedom $(u=1,\ldots,n_0;\,s=1,\ldots,k)$. A large number of outlying standardised prediction errors, with respect to the reference distribution, indicates serious inadequacy of the emulator. \citet{Bastos} suggested various graphical methods for identifying patterns in outliers and, subsequently, causes for emulator inadequacy; for example, plots of the individual prediction errors against each input variable or the predictive mean.

Individual $(1-\alpha) \times 100$\% predictive probability intervals for each element of $Y_0$ can be constructed as

$$q_{us} \pm c_{\alpha} \sqrt{\frac{R_{uu} \hat{S}_{ss}}{\hat{\delta}}}\,,$$

\noindent where $c_{\alpha}$ is the $(1-\alpha/2)$th quantile of the standard t-distribution with $\hat{\delta}$ degrees of freedom. The obtained coverage of these intervals can be compared against $1-\alpha$, with low coverage suggesting the emulator is underestimating the prediction uncertainty.

\subsubsection{Omnibus diagnostic}
\label{OMNIBUS}

We now develop a summary statistic for overall emulator adequacy, analogous to the Mahalanobis distance diagnostic of \citet{Bastos}. Define $E$ as the $n_0 \times k$ matrix of standardised predictions

$$E = G_R^{-1} \left(Y_0 - Q\right)G_S^{-1}\,,$$

\noindent where $R=G_RG_R^\mathrm{T}$ and $\hat{S}=G_S^\mathrm{T} G_S$. Following \citet{Javier}, for an adequate emulator, the conditional posterior distribution of $E$ is

$$E | Y,\mathbf{r} \sim \mathrm{MT}_{n_0,k}\left(0_{n_0\times k},I_k,I_{n_0},\hat{\delta}\right)\,.$$

\noindent We now define the diagnostic
 
\begin{equation}
U = |I_k + E^TE|^{-1}\,,
\label{Ustat}
\end{equation}

\noindent with extreme (large or small) values of $U$, relative to the reference distribution, indicating emulator inadequacy. Following \citet{Dickey}, the reference distribution for $U$ is a $\mathrm{U}_{k,n_0,k+\hat{\delta}-1}$ distribution (conditional on $Y$ and $\mathbf{r}$). \citet[p. 307]{Anderson} showed that the $\mathrm{U}_{k,n_0,k+\hat{\delta}-1}$ distribution has the same distribution as a product of $k$ independent Beta random variables, i.e.

$$\prod_{s=1}^k X_s \sim \mathrm{U}_{k,n_0,k+\hat{\delta}-1},$$

\noindent where $X_s \sim \mathrm{Beta}\left((k+\hat{\delta}-s)/2,n_0/2\right)$. Summaries of this distribution can be calculated by simulation. 

The matrices $G_R$ and $G_S$ are not unique and depend on the chosen decomposition of $R$ and $\hat{S}$, respectively; for example, the eigen or Cholesky decomposition. However,

\begin{eqnarray*}
U & = & | I_k + (G_S^{-1})^\mathrm{T} \left(Y_0 - Q\right)^\mathrm{T} R^{-1}\left(Y_0 - Q\right)G_S^{-1}|^{-1}\\
& = & | I_k + \hat{S}^{-1} \left(Y_0 - Q\right)^\mathrm{T} R^{-1}\left(Y_0 - Q\right)|^{-1}\,,
\end{eqnarray*}

\noindent and therefore the value of the diagnostic $U$ is invariant to the choice of decomposition.

Assuming~\eqref{DISTRIBUTION}, also note that 

$$\mathrm{cov}\left(\mathrm{vec}(E)\right) = \frac{1}{\hat{\delta}-2} I_{kn_0}\,,$$

\noindent and hence the elements of $\hat{\delta}^{\frac{1}{2}}E$ form an uncorrelated sample from the t-distribution with $\hat{\delta}$ degrees of freedom. Quantile-quantile (QQ) plots of these elements can be used as an additional check on emulator adequacy. The elements of $E$ are dependent on the decomposition used to obtain $G_R$ and $G_S$. However as noted by \citet{Bastos}, any choice of decomposition method is appropriate for use in a QQ-plot, and we use the Cholesky decomposition.

For univariate simulator output ($k=1$), the omnibus statistic reverts to the Mahalanobis distance suggested by \citet{Bastos}. Now, $E$ is an $n_0 \times 1$ vector following a $\mathrm{t}_{n_0}(\mathbf{0},(1/\hat{\delta})I_{n_0},\hat{\delta})$ distribution, $E^T E$ is scalar and $1-U \sim \mathrm{Beta}(n_0/2,\hat{\delta}/2)$ with

$$\frac{\hat{\delta} (1-U)}{n_0 U} = \frac{\hat{\delta}}{n_0} E^TE \sim \mathrm{F}\left(n_0,\hat{\delta}\right)\,.$$

\noindent The quantity $E^TE / (\hat{\delta}-2)$ is the Mahalanobis distance and $\mathrm{F}(a,b)$ denotes an $\mathrm{F}$ distribution with $a$ and $b$ degrees of freedom.


\subsection{Emulator improvement}
\label{IMPROVING}

The diagnostics in Section~\ref{DIAGSUB} can be used to suggest improvements to a multivariate emulator. For example, graphical assessment of standardised errors may suggest different mean functions $h(\mathbf{x})$, transformations of inputs, or regions of $\mathcal{X}$ where new simulator runs should be performed; see \citet{Bastos}. We focus on selection of an appropriate mean function and improvement of GP emulators via the addition of a nugget.

\subsubsection{Mean function selection via model comparison} \label{Modelselection}

It is common in the application of GP emulators to usually assume a simple form for the mean function such as $h(\mathbf{x})=1$ or $h(\mathbf{x})=c(1,\mathbf{x})$ (see, for example, \citealp{Bayarri}). Clearly, for the lightweight emulator, with uncorrelated errors, such a simple assumption will usually be inappropriate. We demonstrate in Section~\ref{RESULTS} that using an overly complex mean function (i.e. overfitting) can also be detrimental to the accuracy of the emulator on an independent test data set, as with the more usual applications of the linear model. This motivates the use of Bayesian model comparison as a vehicle for the selection of an appropriate mean function.

Let each unique choice of $h(\mathbf{x})$ be indexed by $v$, i.e. we label mean functions as $h_v(\mathbf{x})$, with $v \in \mathcal{V}$ and $\mathcal{V}$ denoting the set of possible models. Then, following equations~\eqref{DISTRIBUTION} and~\eqref{DISTRIBUTION4},

\begin{equation*}
Y| B_v,\Sigma_v, v, \mathbf{r}_v \sim \mathrm{MN}_{n,k} \left(H_vB_v,\Sigma_v,A_v\right)\,,
\label{MODELDISTRIBUTION}
\end{equation*}

\noindent and

\begin{equation}
Y_0 | Y,v,\mathbf{r}_v \sim \mathrm{MT}_{n_0,k} \left( Q_v, \hat{S}_v, R_v, \hat{\delta}_v \right)\,,
\label{MODELPROB}
\end{equation}

\noindent where

\begin{eqnarray*}
Q_v & = & H_{v,0} \hat{M}_v + T_v^\mathrm{T} A_v^{-1}(Y - H_v\hat{M}_v)\,,\\
R_v & = & A_{v,0} - T_v^\mathrm{T} A_v^{-1} T_v + (H_{v,0} - T_v^\mathrm{T} A_v^{-1}H_v)\hat{\Omega}_v(H_{v,0} - T_v^\mathrm{T} A_v^{-1}H_v)^\mathrm{T}\,,\\
\hat{\Omega}_v & = & \left(H_v^\mathrm{T}A_v^{-1} H_v + \Omega_v^{-1} \right)^{-1}\,,\\
\hat{M}_v & = & \hat{\Omega}_v \left(H_v^\mathrm{T}A_v^{-1} Y + \Omega_v^{-1} M_v \right)\,,\\
\hat{S}_v & = & Y^\mathrm{T}A_v^{-1} Y + M_v^\mathrm{T} \Omega_v^{-1} M_v + S_v - \hat{M}_v^T \hat{\Omega}_v^{-1} \hat{M}_v\,,\\
\hat{\delta}_v & = & \delta_v + n\,,
\end{eqnarray*}

\noindent $M_v$, $\Omega_v$, $S_v$ and $\delta_v$ are hyperparameters for the $v$th model, $\mathbf{r}_v$ holds the correlation parameters for the $v$th model, and $H_{v,0}$, $H_v$, $A_v$, $A_{v,0}$, $T_v$ and $B_v$ for model $v$ are analogous to matrices defined in Section~\ref{METHODOLOGY}.

A fully Bayesian approach would average (\ref{MODELPROB}) with respect to the posterior distribution of the correlation parameters, $\mathbf{r}_v$, and the posterior model probabilities to provide a model-averaged posterior predictive distribution. Alternatively, Bayesian model comparison can be used to identify a model $\hat{v}$, based on the posterior model probabilities, and $Y_0 | Y, \hat{\mathbf{r}}_{\hat{v}},\hat{v}$ can be employed as an emulator. The obvious choice for $\hat{v}$ is the posterior modal model with highest posterior model probability. We adopt this latter approach, both for computational convenience and also to provide interpretable emulators that aid scientific understanding of the simulator.

The posterior model probability for model $v$ is given by

$$\pi(v|Y) = \frac{\pi(v) \int \pi(Y|\mathbf{r}_v,v) \pi(\mathbf{r}_v|v) \mathrm{d}\mathbf{r}_v}{\sum_{v \in \mathcal{V}}\pi(v) \int \pi(Y|\mathbf{r}_v,v) \pi(\mathbf{r}_v|v) \mathrm{d}\mathbf{r}_v}\,,$$

\noindent where $\pi(v)$ is the prior model probability of $v$ such that $\sum_{v \in \mathcal{V}} \pi(v) = 1$,

$$\pi(Y|\mathbf{r}_v,v) = \frac{\Gamma_k\left(\frac{k+\hat{\delta}_v-1}{2}\right)}{\pi^{nk/2}\Gamma_k\left(\frac{k+\delta_v-1}{2}\right)|A_v|^{k/2}}
\frac{|\hat{\Omega}_v|^{k/2}}{|\Omega_v|^{k/2}}\frac{|S_v|^{(\hat{\delta}_v + k -1)/2}}{|\hat{S}_v|^{(\hat{\delta}_v + k -1)/2}}\,,$$

\noindent and $\Gamma_k(\cdot)$ is the multivariate gamma function \citep{Javier}

$$\Gamma_k(x) = \pi^{k(k-1)/4} \prod_{s=1}^k \Gamma\left(x - (s-1)/2\right)\,.$$

\noindent The term $\int \pi(Y|\mathbf{r}_v,v) \pi(\mathbf{r}_v|v) \mathrm{d}\mathbf{r}_v$ which features in the posterior model probability is known as the marginal likelihood. For the GP emulator, the integration required to evaluate the marginal likelihood will not be analytically tractable. For the lightweight emulator, where $A_v=I_n$ and does not depend on $\mathbf{r}_v$, the marginal likelihood is available in closed form. However, if the number of models, $|\mathcal{V}|$, is large then calculating the marginal likelihood for every model will be computationally infeasible. Instead we generate a sample from the posterior distribution of the model index, $v$, using MCMC methods. For a GP emulator, each iteration of the MCMC method has two phases.

\emph{Phase 1} uses the MCMC model composition algorithm \citep{Raftery} to update the model index conditional on the current value of the correlation parameters. Suppose the current model is $v$ and a move to a model $w$ is proposed with probability $\rho(v,w)$ where the correlation parameters remain unchanged, i.e. $\mathbf{r}_w = \mathbf{r}_v$. The move is accepted with probability

\begin{equation}\label{alphaeq}
\alpha = \frac{\pi(Y|\mathbf{r}_v,w) \pi(w)}{\pi(Y|\mathbf{r}_v,v) \pi(v)} \frac{\rho(w,v)}{\rho(v,w)}\,.
\end{equation}

\emph{Phase 2} updates the correlation parameters, $\mathbf{r}_v$, conditional on the current model $v$ using a suitable MCMC method. We employ a random walk Metropolis-Hastings algorithm. 

For the lightweight emulator, phase 2 is not required. After a large number of iterations, when the chain has reached a stationary distribution, the proportion of iterations that visit model $v$ provides an approximation to $\pi(v|Y)$. We choose $\rho(v,w)$ such that (i) proposed models can only add or remove a single term from the current model, adhering to marginality, and (ii) all possible models that obey these conditions are equally likely to be proposed.

\subsubsection{Non-zero nugget}
\label{IMPROVING_MGP}

\citet{Gramacy2} discussed improving the adequacy of univariate GP emulators via the inclusion of a non-zero nugget parameter, principally to mitigate the effects of incorrect model assumptions. Use of a nugget changes the $(i,j)$th element of $A$, 

$$
a_{ij} = c(\mathbf{x}_i,\mathbf{x}_j;\mathbf{r}) + \eta I(i=j)\,,
$$

\noindent where $\eta \ge 0 $ is the nugget parameter and $I(i=j)$ is the indicator function. For prediction, we again adopt a plug-in approach for the nugget parameter, and replace $\eta$ by a representative value $\hat{\eta}$ (the posterior mode). For model selection, the value of the nugget is sampled in phase 2 of the MCMC algorithm. The prior for $\eta$ used in this paper is given by $\pi(\eta) = (1+\eta^2)^{-1}$, previously used by \citet{Conti} for correlation parameters.  


\section{Application to the DIAMOND simulator}
\label{RESULTS}

In this section, the methodology from Sections~\ref{METHODOLOGY} and~\ref{DIAGS} is employed to construct and check multivariate GP and lightweight emulators for the DIAMOND simulator. Recall that the scenario under investigation has been solely designed for model testing purposes. Hence, when, for example, we refer to the importance of specific input variables, we do so only in that context. In particular, we do not intend these observations to be applied to other situations. For the construction of each emulator, we scale the continuous input variables to $[0,1]$ and denote the levels of the categorical variables as $\{0,1\}$.

\subsection{Prior information}
\label{prior}

When constructing individual GP and lightweight emulators, we assume weak prior information for the model parameters $B$, $\Sigma$ and $\mathbf{r}$, following \citet{Conti}:

\begin{eqnarray*}
M & = & 0_{m\times k}\,,\\
\Omega^{-1} & = & 0_{m\times m}\,,\\
S & = & 0_{k\times k}\,,\\
\delta & = & -k+1\,.
\end{eqnarray*}

\noindent The correlation parameters $\mathbf{r}$ are assumed independent, with prior distributions specified using the approach of \citet{Link}. We rewrite $c(\mathbf{x}_1,\mathbf{x}_2;\mathbf{r})$, from~(\ref{CORR_2}), as
 
$$
c(\mathbf{x}_1,\mathbf{x}_2;\mathbf{r}) = \prod_{l=1}^{p_1} \rho_l^{|x_{1l} - x_{2l}|^{2}} \prod_{l=p_1+1}^p \rho_l^{\mathrm{I}(x_{1l} \ne x_{2l})}\,,
$$

\noindent where $\rho_l = \exp (-r_l) \in (0,1)$ for $r_l>0$ ($l=1,\ldots,p$). We assume a uniform prior distribution for $\rho_l$, leading to the induced prior for $r_l$ being an exponential distribution with $E(r_l)=1$.

When performing model comparison for the selection of the mean function with only weak prior information available for the parameters of each model, we adopt prior hyperparameters $S_v=0_{k\times k}$ and $\delta_v=-k+1$ for $\Sigma_v$, which is present in all models, and unit information prior distributions for $B_v$, with $M_v=0_{p\times p}$ and

$$\Omega_v = n \left(H_v^\mathrm{T} A_v^{-1} H_v\right)^{-1}\,,$$

\noindent as proposed by \citet{Kass}. The use of proper prior distributions for $B_v$ avoids Lindley's paradox \citep[see][pg 394]{Bernardo} which states that the posterior model probabilities are sensitive to the scale of the prior variance (see also \citealp[pp. 322-324]{OHagan}, \citealp{Raftery} and \citealp{Fernandez}). We assume the same exponential prior, see above, for each element of $\mathbf{r}_v$ for each model, i.e. $\pi(\mathbf{r}_v|v) = \pi(\mathbf{r}_v)$. A uniform prior over the model space is chosen, i.e. $\pi(v)=|\mathcal{V}|^{-1}$, where $\mathcal{V}$ is the set of all sub-models of the maximal model that respect marginality. The maximal model has a mean function consisting of the intercept, all linear, two-way interaction and, for the continuous inputs, quadratic terms. The resulting model matrix, $H$, has $m=103$ columns.


For this weak prior information, $\alpha$ from~\eqref{alphaeq} reduces to

$$\alpha = (n+1)^{k(m_v-m_w)/2} \frac{|\hat{S}_v|^{n/2}}{|\hat{S}_w|^{n/2}}\frac{\rho(w,v)}{\rho(v,w)}\,,$$

\noindent where

$$\hat{S}_v = Y^\mathrm{T} A_v^{-1} \left( I_n - \frac{n}{n+1} H_v \left(H_v^{\mathrm{T}} A_v^{-1} H_v \right)^{-1} H_v^\mathrm{T}A_v^{-1} \right) Y\,.$$


\subsection{Design of the computer experiment}
\label{DESIGN}

We employed a space-filling design that would enable the estimation of both the Gaussian process and lightweight emulators. The most common design used for computer experiments is the Latin Hypercube \citep{McKay} and its extensions (see, for example, \citealp{Tang}, and \citealp{Morris}). Such designs provide low-dimensional uniformity in the input variables, hence achieving good projection properties,  and allow the estimation of nonparametric regression models. They are also an attractive choice for lightweight emulation, as the exact form of the emulator will be unknown in advance of the data collection and a flexible design that allows the fitting of many different parametric models may be required (see Section~\ref{IMPROVING}).

The design, $\zeta = \left\{\mathbf{x}_1,...,\mathbf{x}_n \right\}$, for this study needed to combine both continuous and categorical input variables. We used a sliced space-filling design as proposed by \citet{Qian2} with $n=120$ runs. Such a design, constructed from an orthogonal array, has not only good space-filling properties overall but also for the projection into the continuous variables for each combination of values of the categorical input variables.

\subsection{Construction of adequate emulators}
\label{EMCONSTR}

We constructed both lightweight and multivariate GP emulators for the DIAMOND simulator using the $n=120$ simulator runs, each outputting $k=5$ responses, from the sliced space-filling design as training data. For model validation and diagnostics, we use a second design $\zeta_0= \left\{\mathbf{x}_{01},...,\mathbf{x}_{0n_0} \right\}$, with associated $n_0\times k$ simulator output matrix $Y_0$. This design is also a sliced space-filling design with $n_0=120$ runs and was constructed using a different orthogonal array to that used to construct $\zeta$.

We chose, assessed and compared emulators using the diagnostics from Section~\ref{DIAGS}. We calculated the root mean squared error (RMSE) for $Y_0$,


$$\mathrm{RMSE} = \left[ \frac{1}{n_0k} \sum_{u=1}^{n_0} \sum_{s=1}^k \left(Y_{0,us} - q_{us} \right)^2 \right]^{1/2}\,,$$

\noindent where $Y_{0,us}$ is the simulator output from the $u$th validation run for response $s$. We also calculated the root relative mean squared error (RRMSE),

$$\mathrm{RRMSE} = \left[ \frac{1}{n_0k} \sum_{u=1}^{n_0} \sum_{s=1}^k \frac{\left(Y_{0,us} - \gamma_{us} \right)^2}{Y_{0,us}^2} \right]^{1/2},$$

\noindent where the point estimate $\gamma_{us} = \mathrm{E}\left(Y_{0,us}^{-1}|Y,\hat{\mathbf{r}}\right)/\mathrm{E}\left(Y_{0,us}^{-2}|Y,\hat{\mathbf{r}}\right)$ minimises the relative squared error loss function.

\subsubsection{Lightweight emulators \label{LWEres}}
\begin{figure}
\centering
\includegraphics[scale=0.7, viewport = 50 423 545 670, clip=true]{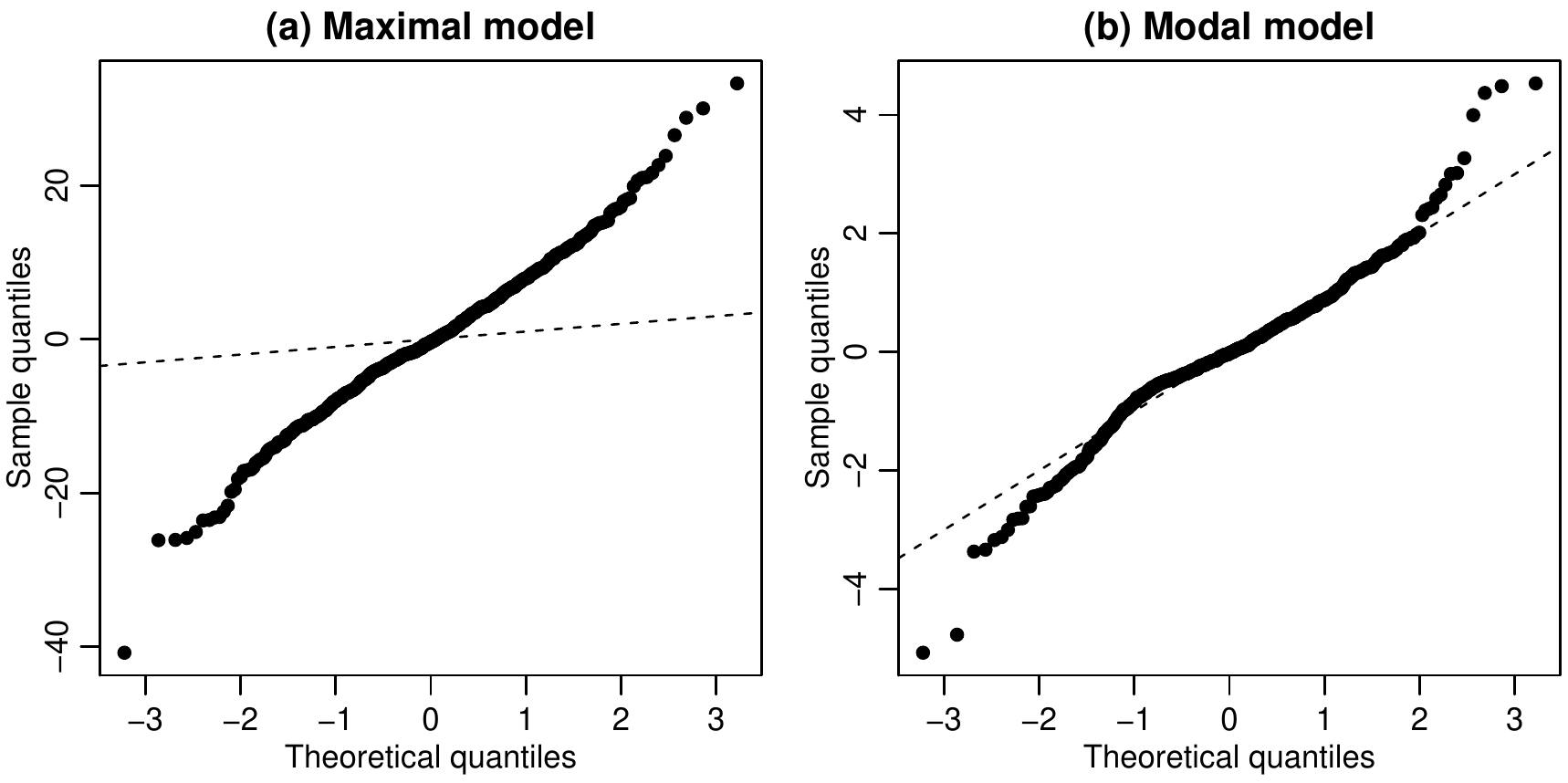}
\caption{\label{LWE_QQ}QQ-plots of the uncorrelated errors against a reference t-distribution for lightweight emulators: (a) the maximal model; (b) the modal model}
\end{figure}

Our first lightweight emulator was the maximal model. The value of the omnibus test statistic, $U$, and coverage of the 95\% predictive probability intervals are given in Table~\ref{VALUES}. Note that the reference distribution for $U$ has expected value of 0.030, and 2.5\% and 97.5\% quantiles of 0.019 and 0.044, respectively. The diagnostics indicate there is a discrepancy between the simulator and this emulator, with the observed value of $U$ and the achieved coverage both being low. Further evidence of this discrepancy is the QQ-plot of the uncorrelated errors against a reference t-distribution, Figure~\ref{LWE_QQ}(a); the points form a line with slope greater than one, indicating that the variance associated with the emulator predictions has been underestimated.

\begin{table}
\caption{Observed values (to 3 decimal places) of the omnibus diagnostic $U$, coverage of the 95\% predictive probability intervals, RMSE and RRMSE for the various emulators considered.
The reference distribution for $U$ has expected value of 0.030, and 2.5\% and 97.5\% quantiles of 0.019 and 0.044, respectively  \label{VALUES}}
\centering
\begin{tabular}{lllllll} \hline
Emulator & Mean function & Nugget & $U$   & Coverage & RMSE    & RRMSE \\ \hline
Lightweight      & Maximal       & NA     & 0.000 & 0.478    & 2728.791 & 6.975 \\
      & Modal         & NA     & 0.025 & 0.953    & 988.729 & 0.528 \\ \hline
Multivariate GP      & Intercept     & Zero   & 0.001 & 0.958    & 415.030 & 0.457 \\
      & Linear        & Zero   & 0.015 & 0.965    & 344.234 & 0.397 \\
      & Modal         & Zero   & 0.012 & 0.958    & 341.859 & 0.396 \\
			& Maximal 			& Zero 	 & 0.000 & 0.477    & 2701.149 & 6.791 \\ \hline
Multivariate GP      & Intercept     & Non-zero   & 0.033 & 0.975    & 363.014 & 0.387 \\
      & Linear        & Non-zero   & 0.019 & 0.948    & 1264.094 & 0.539 \\
      & Modal         & Non-zero   & 0.034 & 0.963    & 334.597 & 0.403 \\
			& Maximal 			& Non-zero 	 & 0.000 & 0.478    & 2728.383 & 6.973 \\ \hline
\end{tabular}
\end{table}

To attempt to alleviate the obvious inadequacy of this emulator, alternative mean functions $h(\mathbf{x})$ were compared using Bayesian model comparison (Section~\ref{IMPROVING}). The posterior modal model was found from $10^5$ iterations of the MCMC algorithm (discarding the first 10\% of iterations as burn in). The algorithm took 2.5 minutes on a computer with a 3.20Ghz processor and 8Gb RAM, and the average acceptance rate for the proposed moves in Phase 1 was 4.7\%, reflecting the concentration of the posterior model probabilities on a small number of models. Table~\ref{MARG_PROB} displays the terms in the posterior modal model, and gives the associated posterior marginal inclusion probabilities (i.e. the proportion of visited models that included that term). The model matrix, $H$, for the posterior modal model has $m=11$ columns. The value of $U$ and the coverage for the emulator with this alternative mean function are shown in Table~\ref{VALUES}. These values suggest there is no evidence of a discrepancy between the simulator and the emulator. This conclusion is supported by the QQ-plot of the uncorrelated errors in Figure~\ref{LWE_QQ}(b). Also shown in Table~\ref{VALUES} are the RMSE and the RRMSE of the maximal and modal model emulators. Note how the simpler form of emulator has smaller values for RMSE and RRMSE, indicating the modal model has significantly improved predictive accuracy.

\subsubsection{Multivariate Gaussian process emulators}

We construct GP emulators with four different forms for the mean function, $h(\mathbf{x})$: (i) intercept only ($m=1$); (ii) linear terms only ($m=8$); (iii) the modal model found by the model comparison procedure ($m=7$; see Table~\ref{MARG_PROB}); and (iv) the maximal model ($m=103$). We initially fix the nugget at zero. As a comparison with Section~\ref{LWEres}, the model comparison procedure took 30 minutes and had an acceptance rate of 0.5\%.

\begin{figure}[t]
\centering
\includegraphics[scale=0.7, viewport = 45 420 551 669, clip=true]{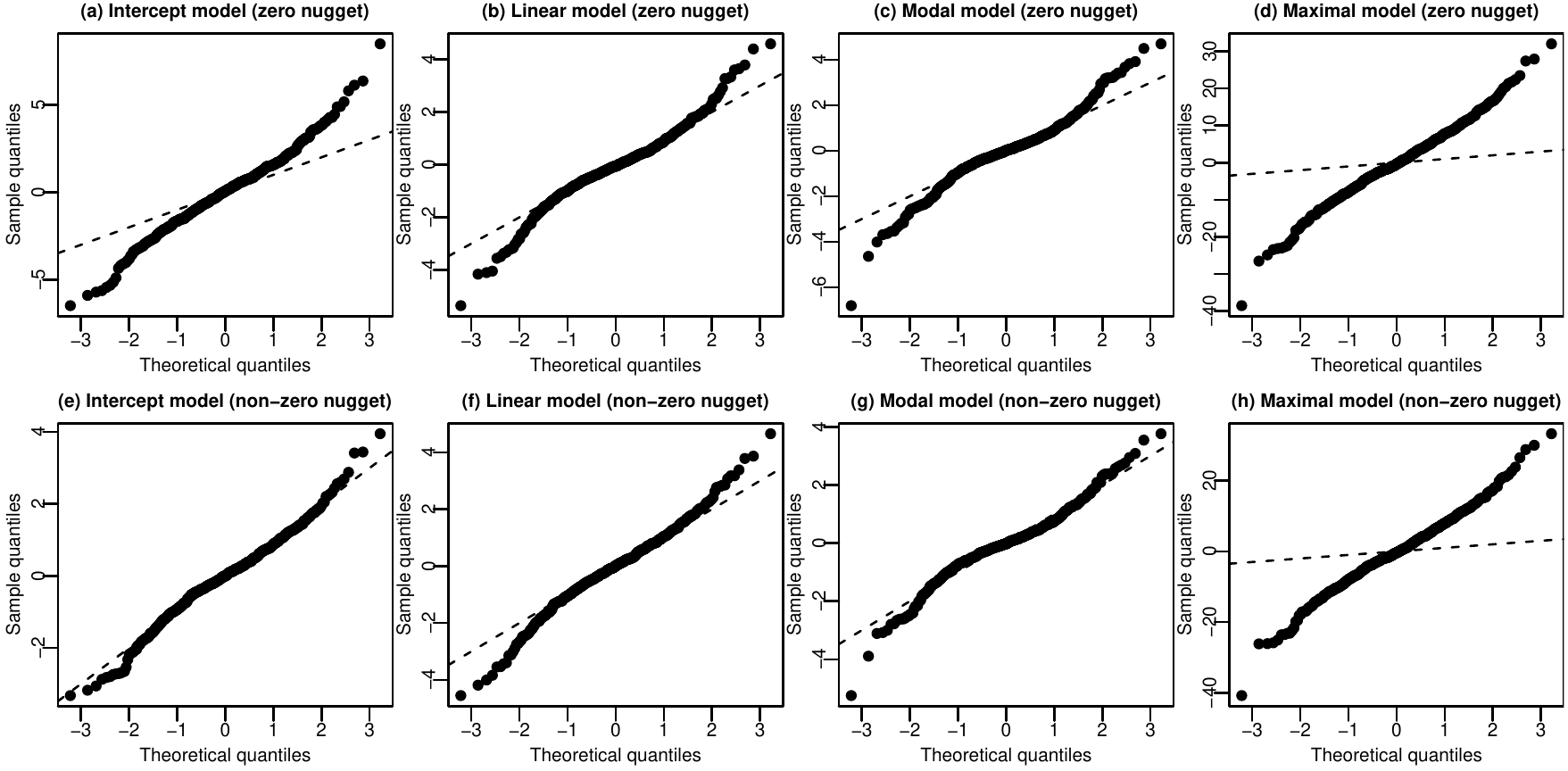}
\caption{\label{MO_plot}QQ-plots of the uncorrelated errors against a reference t-distribution for the zero nugget GP emulator with (a) the intercept model, (b) the linear model, (c) the modal model, (d) the maximal model, and for the non-zero nugget GP emulator with (e) the intercept model, (f) the linear model, (g) the modal model, (h) the maximal model}
\end{figure}

Table~\ref{VALUES} shows the values of $U$ and the coverage for these four GP emulators. Figure~\ref{MO_plot}(a-d) show QQ-plots of the uncorrelated errors for these emulators. Clearly, the values in Table~\ref{VALUES} and the QQ-plots show that there exist serious discrepancies between all four emulators and the simulator. Similar to the maximal lightweight emulator, the QQ-plot shows that the variances associated with the GP emulator predictions are underestimated.


\begin{table}
\caption{\label{MARG_PROB}Marginal posterior probabilities (up to 3 decimal places) of the terms in the modal mean functions}
\centering
\begin{tabular}{lcrrr} \hline
Terms && Lightweight & GP & GP \\ \
 && & (zero & (non-zero \\
 && & nugget) & nugget) \\ \hline
Linear Effects && & & \\ \
Food capacity (Giarre) & $x_3$ & 0.999 & 1.000 & 1.000 \\ \
Food capacity (Catania) & $x_6$ & 1.000 & 1.000 & 1.000 \\ \
Planning time & $x_8$ & 0.970 & 1.000 & 1.000 \\ \
Recipient of food aid & $x_{12}$ & 1.000 & - & - \\ \
Location of NGO base & $x_{13}$ & 1.000 & 1.000 & 1.000 \\ \hline
Quadratic Effects && & & \\ \
Planning time && 0.764 & 0.999 & 1.000 \\ \hline
Interactions && & & \\ \
Food capacity (Giarre) $\times$ Recipient of food aid && 0.811 & - & - \\ \
Food capacity (Catania) $\times$ Recipient of food aid && 1.000 & - & - \\ \
Food capacity (Catania) $\times$ Location of NGO base && 1.000 & 0.983 & 0.828 \\ \
Recipient of food aid $\times$ Location of NGO base && 0.914 & - & - \\ \hline
\end{tabular}
\end{table}

To remedy these inadequacies, we included a non-zero nugget  in emulators using all four forms of the mean function. The model comparison algorithm took 30 minutes and had an acceptance rate of 1.6\%. The modal mean function for both types of GP emulator (with and without nugget) are identical (see Table~\ref{MARG_PROB}). The values of $U$ and the coverage for the four non-zero nugget GP emulators are also shown in Table~\ref{VALUES}. The corresponding QQ-plots are shown in Figure~\ref{MO_plot}(e-h). There still exist discrepancies between the emulator and simulator for the maximal and linear forms of the mean function. However, for the intercept and modal forms, the values in Table~\ref{VALUES} and the QQ-plots provide no evidence of inadequacy, with the diagnostics being highly plausible under their reference distributions. The values of RMSE and RRMSE for all eight GP emulators are also given in Table~\ref{VALUES}. Note the high values of these errors under the maximal models. The intercept and modal GP emulators (with non-zero nugget) have significantly higher predictive accuracy than the lightweight emulators. There appears to be little difference between the intercept and modal model for the GP emulators (with non-zero nugget) in terms of predictive accuracy.

\subsection{Sensitivity analysis}

An important application of statistical emulators are sensitivity analyses to identify important input variables and their impact on the responses. For the lightweight emulator, the model comparison algorithm in Section~\ref{Modelselection} has the advantage of automatically identifying the most important input variables. When product terms are included in the mean function, it can also identify important interactions. For the DIAMOND simulator, there are interactions between the food capacity at Catania and both the location of the NGO base and the recipient of the food aid. There are also interactions between the food capacity at Giarre and the recipient of food aid and location of NGO base and recipient of food aid. There is evidence that planning time has a non-linear effect.


\begin{figure}[!t]
\centering
\includegraphics[scale=0.6, viewport = 50 173 533 669, clip=true]{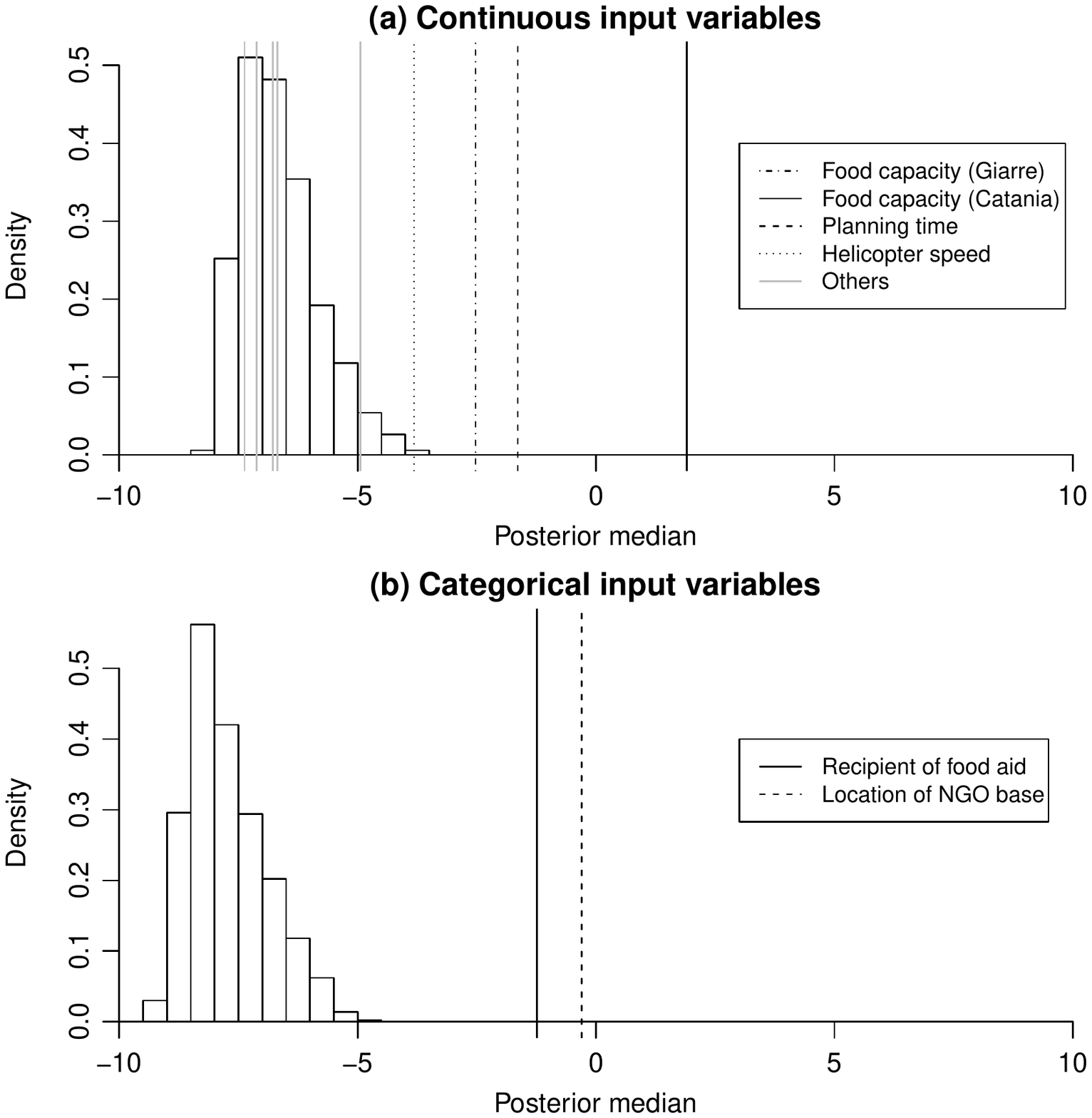}
\caption{\label{HISTO}Histograms of the null reference distributions from the RDVS method for the correlation parameters of the (a) continuous and (b) categorical inert input variables. The posterior medians of the input variables are shown as vertical lines}
\end{figure}

For the multivariate GP emulator, input variables can impact the response through both the mean function and correlation structure. Hence, the model selection algorithm in Section~\ref{Modelselection} may not identify all the important variables. For an intercept-only GP, the relative importance of the input variables is only determined by the relative magnitude of the corresponding correlation parameters, $\mathbf{r}$.  In general, the output is more sensitive to those input variables with large correlation parameters. As calibrating the size of correlation parameters can be difficult, \citet{Link} proposed a more formal variable selection method for univariate GPs, reference distribution variable selection (RDVS). Values of an inert input variable, $x^*$, are randomly generated from the input space $\mathcal{X}$. An MCMC sample is generated from the marginal posterior distribution of $\mathbf{r}$ and $r^*$, where $r^*$ is the correlation parameter of the inert input variable. The above procedure is repeated $B$ times with different randomly generated values of inert input variables. The posterior median of each element of $\mathbf{r}$, approximated from the union of the MCMC samples from all randomly generated sets of inert input variables, is compared to the null reference distribution of the posterior medians of $r^*$ (obtained from the $B$ sets of values for the inert input variable). For more details see \citet{Link}.

Application of RDVS to multivariate GP emulators is straightforward. Our simulator has both continuous and categorical input variables, and hence we adapt RDVS by at each iteration randomly generating values for two inert input variables, $x^*_1$ and $x^*_2$, where $x^*_1\in[0,1]$ and $x^*_2\in\{0,1\}$, with $\{0,1\}$ indicating the two levels for a categorical variable. The posterior median of the elements of $\mathbf{r}$ corresponding to continuous input variables is then compared to the null reference distribution of the posterior medians of $r^*_1$, and similarly for the categorical input variables and $r_2^*$.

We applied RDVS with the GP emulator (with non-zero nugget and the intercept mean function), using $B=1000$. Figure~\ref{HISTO} displays the null reference distributions for the correlation parameters (on the log scale) of the (a) continuous and (b) categorical inert input variables, i.e. the 1000 posterior medians of the correlation parameters, $r_1^*$ and $r_2^*$, from the MCMC samples. Also indicated in Figure~\ref{HISTO} are the posterior medians of the actual input variables as vertical lines. Clearly the most important continuous input variables are the food capacities at both Giarre and Catania, planning time and helicopter speed. Both of the categorical input variables are deemed to be important. This agrees with the conclusions from the modal lightweight emulator, except for the inclusion of helicopter speed. 

RDVS with a GP emulator having mean function including only an intercept is unable to explicitly identify interactions. A probabilistic sensitivity analysis (see, for example, \citealp{Santner}, ch. 7) can be used to understand and visualise the functional form of the individual and joint effects of the variables. 

The variation in the simulator output induced by variation in the input variables can be decomposed into main effects and interactions. Assume interest is in the total number of casualties across days two to six of the disaster, given by $g(\mathbf{x}) = \sum_{i=1}^k f_i(\mathbf{x})$. Letting $\mathrm{E}$ denote expectation with respect to an assumed joint distribution for the input variables $\mathbf{x}$, we can then define the following main effects and first-order interactions:
\begin{eqnarray}
g_i(x_i) & = & \mathrm{E}\left[g(\mathbf{x})|x_i\right] - g_0\,, \label{maineffect}\\
g_{ij}(x_i,x_j) & = & \mathrm{E}\left[g(\mathbf{x})|x_i,x_j\right] - g_0 - g_i(x_i) - g_j(x_j)\,,\label{inteffect}
\end{eqnarray}
where $g_0 = \mathrm{E}\left[g(\mathbf{x})\right]$. Corresponding partial variances are given by
\begin{eqnarray*}
V_i & = & \mathrm{E}\left[g_i(x_i)^2\right]\,,\\
V_{ij} & = & \mathrm{E}\left[g_{ij}(x_i,x_j)^2\right]\,,\qquad i,j=1,\ldots,p\,. 
\end{eqnarray*}
Following \citet{Oakley}, these variances can be estimated by their expectation, denoted $\mathrm{E}^*$, with respect to the posterior predictive distribution of $g(\mathbf{x})$, a non-standard $t$ distribution; see Section~5 of the Supplementary Material. Hence, the following estimated sensitivity indices can be defined:
\begin{align*}
\hat{\mathcal{S}}_i & =  \mathrm{E}^*(V_i)/\mathrm{E}^*(V)\,, & \mbox{(first-order)}\\
\hat{\mathcal{S}}_{ij} & =  \mathrm{E}^*(V_{ij})/\mathrm{E}^*(V)\,, & \mbox{(second-order)}
\end{align*}
where $V = \mbox{Var}\left[g(\mathbf{x})\right]$ with respect to the distribution of the input variables. Explicit formulae for $\mathrm{E}^*(V)$, $\mathrm{E}^*(V_i)$ and $\mathrm{E}^*(V_{ij})$ can be derived in terms of the expectation with respect to the distribution of the input variables, and are given in Section~6 of the Supplementary Material.

We assume that the input variables are independent, that the continuous variables are uniformly distributed over their corresponding ranges and the categorical input variables have probability $0.5$ on each of their two levels. We compute the estimated sensitivity indices under both the multivariate GP emulator (intercept mean function and non-zero nugget) and, for comparison, the lightweight emulator (modal mean function). For the lightweight emulator, the estimated sensitivity indices are available in closed-form \citep{Rougier} and can only be non-zero for those main effects and interactions featuring in the, selected, modal model. Under the multivariate GP emulator, the expectations with respect to the distribution of the input variables require approximation, achieved here using Monte Carlo integration.

\begin{table}
\caption{\label{sobols} Estimated first- and second-order sensitivity indices (multiplied by 1000 and displayed up to 3 decimal places) of the input variables under the lightweight and multivariate Gaussian process emulators.}
\centering
\begin{tabular}{lcrr} \hline
Terms && Lightweight & Multivariate GP \\ \hline
First-order && &  \\ \
Food capacity (Giarre) & $x_{3}$ & 9.978 & 7.854 \\ \
Food capacity (Catania) & $x_6$ & 887.818 & 895.176 \\ \
Planning time & $x_8$ & 2.589 & 1.881 \\ \
Helicopter speed & $x_9$ & 0.000 & 0.312 \\ \
Recipient of food aid & $x_{12}$ & 2.566 & 2.264 \\ \
Location of NGO base & $x_{13}$ & 63.739 & 64.067 \\ \
Sum of Others && 0.000 & 0.023 \\ \hline
Second-order && &  \\ \
Food capacity (Giarre) $\times$ Recipient of food aid && 1.184 & 0.474 \\ \
Food capacity (Giarre) $\times$ Location of NGO base && 0.000 & 0.365 \\ \
Food capacity (Catania) $\times$ Recipient of food aid && 1.620 & 2.121 \\ \
Food capacity (Catania) $\times$ Location of NGO base && 3.599 & 6.750 \\ \
Planning time $\times$ Location of NGO base && 0.000 & 0.572 \\ \
Planning time $\times$ Food capacity (Catania) && 0.000 & 0.173 \\ \
Recipient of food aid $\times$ Location of NGO base && 1.099 & 0.906 \\ \
Sum of Others && 0.000 & 0.178 \\ \hline
\end{tabular}
\end{table}

\begin{figure}
\centering
\includegraphics[scale=0.8, trim={50 200 0 0}, clip=true]{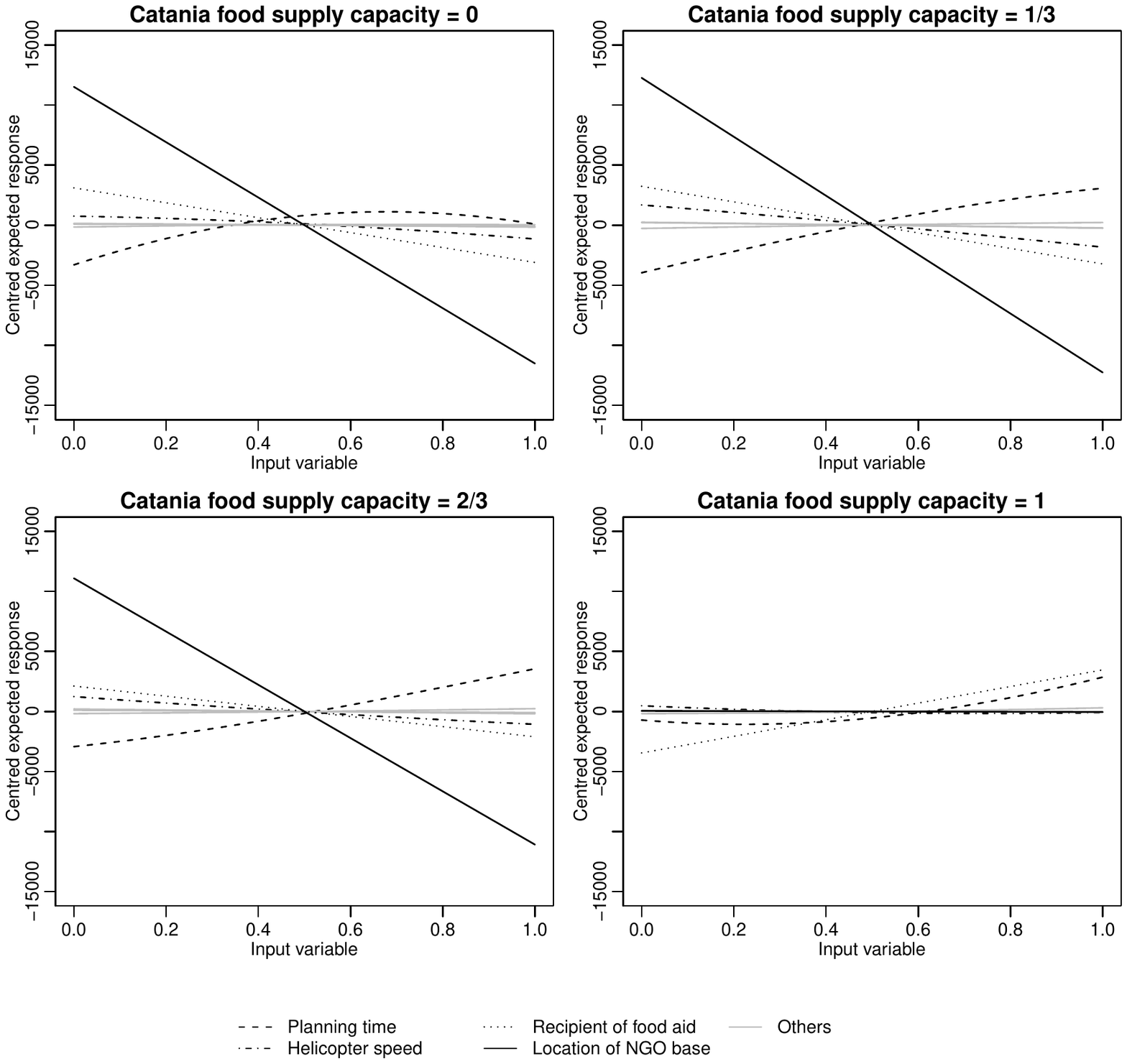}
\caption{\label{plan_plot}Plots of expected conditional main effects~\eqref{condME} from the multivariate GP emulator (intercept mean function and non-zero nugget) for four different settings for the food capacity at Catania ($x_6$).}
\end{figure}

Table~\ref{sobols} shows the estimated sensitivity indices under both emulators. For the multivariate GP, we present first-order estimated sensitivity indices for each of the variables identified by the RDVS method. We also present the seven largest second-order sensitivity indices; four of the corresponding interactions were selected in the modal lightweight emulator. The dominance of input variable $x_6$, controlling the food capacity at Catania, is clear; variation in $x_6$ induces nearly 90\% of the total output variation for both emulators. However, this input variable, in common with $x_1$-$x_5$ is essentially a noise variable and clearly could not be controlled in a real disaster. Hence, of particular interest are the interactions between $x_6$ and the control variables $x_7$-$x_{13}$. To graphically investigate these effects for the GP emulator, in Figure~\ref{plan_plot} we display the expected conditional main effects
\begin{equation}\label{condME}
\mathrm{E}^*\left\{\mathrm{E}\left[g(\mathbf{x})|x_i,x_6=l\right]-g_0\right\}\,,
\end{equation}
for $i=8,9,12,13$ (as identified by RDVS) and $l=0,1/3,2/3,1$. For $x_6\ne 1$, there are strong negative conditional effects for both categorical variables $x_{12}$ and $x_{13}$, with lower casualties resulting from providing food aid only to Catania and, especially, locating the NGO base with the task force. However, for $x_6=1$, variable $x_{13}$ no longer has a substantive effect and $x_{12}$ now has a positive effect (lower casualties result from providing food aid to both cities). Planning time ($x_8$) always has a positive effect, although the degree of nonlinearity changes with the value of $x_6$

\section{Discussion}\label{DISC}

Statistical emulation of multivariate simulators is an important problem in a number of application areas and presents challenging methodological issues. We have presented a unified Bayesian approach to the construction of both parametric (lightweight, linear model) and nonparametric (Gaussian process) emulators, including model selection, diagnostics and sensitivity analyses. Our application, emulating a humanitarian relief simulator applied to an artificial scenario involving an earthquake and volcanic eruption in Sicily, demonstrated the utility and versatility of the methodology. We were able to identify the most important input variables, and their interactions, using the lightweight emulator. While the GP emulator was more accurate, the lightweight emulator was more scientifically intuitive and informative. The technology in this paper provides the capacity for our collaborators to efficiently explore ``what-if'' questions and to make faster ``in-theatre'' decisions.  

Extensions of the methodology to allow the construction and model-checking of different emulators are possible. In Section~\ref{RESULTS}, only weakly informative prior distributions were assumed. If more informative prior information was available, this could be incorporated into both lightweight and GP emulators, for example via the prior distribution for the regression parameters $B|\Sigma$. It is likely that the use of such information would lead to a smaller difference in predictive accuracy between the two emulators, provided there was not a conflict between the prior information and the simulator.

Diagnostics for multivariate emulators were also employed by \citet{Fricker} in a number of case studies using models with a general class of non-separable covariance structure. These diagnostics were similar in spirit to those of \citet{Bastos} but, for example, the non-separability prevents analytic marginalisation across any of the scale parameters when calculating the equivalent to the omnibus statistic~\eqref{Ustat}. An alternative non-separable model may be constructed as the full posterior distribution under model uncertainty, see Section~\ref{Modelselection}. The model-averaged posterior predictive distribution is then a mixture of matrix t-distributions; see also \citet{Rougier}, who proposed a mixture of matrix normal inverse Wishart joint prior distributions for $B$ and $\Sigma$. The diagnostics described in Section~\ref{DIAGSUB} are straightforward to extend to mixture distributions by averaging over the components of the mixture using simulation.

 

\section*{Acknowledgments}\label{ack}
This work was funded by a US Defence Threat Reduction Agency basic research grant and the Defence Science and Technology Laboratory (Dstl). D. C. Woods was supported by Fellowship EP/J018317/1 from the UK Engineering and Physical Sciences Research Council. The authors wish to thank Robin Ashmore from Dstl for providing the simulator and data, and related invaluable conversations. The paper was improved by helpful comments from an Associate Editor and two referees.

\bibliographystyle{asa}
\bibliography{sicily}

\begin{thebibliography}{33}
\newcommand{\enquote}[1]{``#1''}
\expandafter\ifx\csname natexlab\endcsname\relax\def\natexlab#1{#1}\fi

\bibitem[{Anderson(2003)}]{Anderson}
Anderson, T.~W. (2003), \textit{An Introduction to Multivariate Statistical
  Analysis}, New York: Wiley, 3rd ed.

\bibitem[{Bastos and {O'Hagan}(2009)}]{Bastos}
Bastos, L.~S. and {O'Hagan}, A. (2009), \enquote{Diagnostics for {G}aussian
  process emulators,} \textit{Technometrics}, 51, 425--438.

\bibitem[{Bayarri et~al.(2007)Bayarri, Berger, Paulo, Sacks, Cafeo, Cavendish,
  Lin, and Tu}]{Bayarri}
Bayarri, M.~J., Berger, J.~O., Paulo, R., Sacks, J., Cafeo, J.~A., Cavendish,
  J., Lin, C., and Tu, J. (2007), \enquote{A framework for validation of
  computer models,} \textit{Technometrics}, 49, 138--154.

\bibitem[{Bernardo and Smith(1994)}]{Bernardo}
Bernardo, J.~M. and Smith, A. F.~M. (1994), \textit{Bayesian Theory},
  Chichester: Wiley.

\bibitem[{Conti and {O'Hagan}(2010)}]{Conti}
Conti, S. and {O'Hagan}, A. (2010), \enquote{Bayesian emulation of complex
  multi-output and dynamic computer models,} \textit{Journal of Statistical
  Planning and Inference}, 140, 640--651.

\bibitem[{Dawid(1981)}]{Dawid}
Dawid, A.~P. (1981), \enquote{Some matrix-variate distribution theory:
  notational considerations and a {B}ayesian application,} \textit{Biometrika},
  68, 265--274.

\bibitem[{Dickey(1967)}]{Dickey}
Dickey, J.~M. (1967), \enquote{Matricvariate generalisations of the
  multivariate t-distribution and the inverted multivariate t-distribution,}
  \textit{Annals of Mathematical Statistics}, 38, 511--518.

\bibitem[{Fang et~al.(2006)Fang, Li, and Sudjianto}]{fls}
Fang, K., Li, R., and Sudjianto, A. (2006), \textit{Design and Modelling for
  Computer Experiments}, Boca Raton: Chapman and Hall.

\bibitem[{Fernandez et~al.(2001)Fernandez, Ley, and Steel}]{Fernandez}
Fernandez, C., Ley, E., and Steel, M. F.~J. (2001), \enquote{Benchmark priors
  for {B}ayesian model averaging,} \textit{Journal of Econometrics}, 100,
  381--427.

\bibitem[{Fricker et~al.(2013)Fricker, Oakley, and Urban}]{Fricker}
Fricker, T.~E., Oakley, J.~E., and Urban, N.~M. (2013), \enquote{Multivariate
  emulators with nonseparable covariance structures,} \textit{Technometrics},
  55, 47--56.

\bibitem[{Gramacy and Lee(2012)}]{Gramacy2}
Gramacy, R.~B. and Lee, H. K.~H. (2012), \enquote{Cases for the nugget in
  modeling computer experiments,} \textit{Statistics and Computing}, 22.

\bibitem[{Guidoboni et~al.(2007)Guidoboni, Ferrari, Mariotti, Comastri,
  Tarabusi, and Valensise}]{cfti}
Guidoboni, E., Ferrari, G., Mariotti, D., Comastri, A., Tarabusi, G., and
  Valensise, G. (2007), \enquote{{CFTI4Med}: Catalogue of strong earthquakes in
  {Italy} (461 {B.C.} - 1997) and the {Mediterranean} area (760 {B.C.} -
  1500).} .

\bibitem[{Ingber et~al.(1991)Ingber, Fujio, and Wehner}]{ingber}
Ingber, L., Fujio, H., and Wehner, M.~F. (1991), \enquote{Mathematical
  comparison of combat computer models to exercise data,} \textit{Mathematical
  Computer Modelling}, 15, 65--90.

\bibitem[{Javier and Gupta(1985)}]{Javier}
Javier, W.~R. and Gupta, A.~K. (1985), \enquote{On matric variate t
  distributions,} \textit{Communications in Statistics - Theory and Methods},
  14, 1413--1425.

\bibitem[{Kass and Wasserman(1995)}]{Kass}
Kass, R.~E. and Wasserman, L. (1995), \enquote{A reference {B}ayesian test for
  nested hypotheses and its relationship to the {S}chwarz criterion,}
  \textit{Journal of the American Statistical Association}, 90, 928--934.

\bibitem[{Kennedy et~al.(2006)Kennedy, Anderson, Conti, and
  {O'Hagan}}]{kaco2006}
Kennedy, M.~C., Anderson, C.~W., Conti, S., and {O'Hagan}, A. (2006),
  \enquote{Case studies in {G}aussian process modelling of computer codes,}
  \textit{Reliability Engineering and System Safety}, 91, 1301--1309.

\bibitem[{Kennedy and {O'Hagan}(2001)}]{Kennedy}
Kennedy, M.~C. and {O'Hagan}, A. (2001), \enquote{Bayesian calibration of
  computer models (with discussion),} \textit{Journal of the Royal Statistical
  Society, {B}}, 63, 425--464.

\bibitem[{Levy and Steinberg(2010)}]{LevySteinberg2010}
Levy, S. and Steinberg, D.~M. (2010), \enquote{Computer experiments: a review,}
  \textit{AStA Advances in Statistical Analysis}, 4, 311--324.

\bibitem[{Linkletter et~al.(2006)Linkletter, Bingham, Hengartner, and
  Ye}]{Link}
Linkletter, C., Bingham, D., Hengartner, N., and Ye, K. (2006),
  \enquote{Variable selection of {G}aussian process models in computer
  experiments,} \textit{Technometrics}, 48, 478--490.

\bibitem[{Mc{K}ay et~al.(1979)Mc{K}ay, Beckman, and Conover}]{McKay}
Mc{K}ay, M.~D., Beckman, R.~J., and Conover, W.~J. (1979), \enquote{A
  comparison of three methods for selecting values of input variables in the
  analysis of output from a computer code,} \textit{Technometrics}, 21,
  239--245.

\bibitem[{Morris and Mitchell(1995)}]{Morris}
Morris, M.~D. and Mitchell, T.~J. (1995), \enquote{Exploratory designs for
  computer experiments,} \textit{Journal of Statistical Planning and
  Inference}, 43, 381--402.

\bibitem[{Oakley and {O'Hagan}(2004)}]{Oakley}
Oakley, J.~E. and {O'Hagan}, A. (2004), \enquote{Probabilistic sensitivity
  analysis of complex models: a {B}ayesian approach,} \textit{Journal of the
  Royal Statistical Society, {B}}, 66, 751--769.

\bibitem[{O'Hagan(2006)}]{ohagan2006}
O'Hagan, A. (2006), \enquote{Bayesian analysis of computer code outputs: A
  tutorial,} \textit{Reliability Engineering and System Safety}, 91,
  1290--1300.

\bibitem[{O'Hagan and Forster(2004)}]{OHagan}
O'Hagan, A. and Forster, J.~J. (2004), \textit{Bayesian Inference}, vol.~2B of
  \textit{Kendall's Advanced Theory of Statistics}, London: Arnold, 2nd ed.

\bibitem[{Qian and Wu(2009)}]{Qian2}
Qian, P. Z.~G. and Wu, C. F.~J. (2009), \enquote{Sliced space-filling designs,}
  \textit{Biometrika}, 96, 945--956.

\bibitem[{Qian et~al.(2008)Qian, Wu, and Wu}]{Qian1}
Qian, P. Z.~G., Wu, H., and Wu, C. F.~J. (2008), \enquote{Gaussian Process
  Models for Computer Experiments with Qualitative and Quantitative Factors,}
  \textit{Technometrics}, 50, 383--396.

\bibitem[{Raftery et~al.(1997)Raftery, Madigan, and Hoeting}]{Raftery}
Raftery, A.~E., Madigan, D., and Hoeting, J.~A. (1997), \enquote{Bayesian model
  averaging for linear regression models,} \textit{Journal of the American
  Statistical Association}, 92, 179--191.

\bibitem[{Rasmussen and Williams(2006)}]{rasmussen}
Rasmussen, C.~E. and Williams, C. K.~I. (2006), \textit{Gaussian processes for
  machine learning}, Cambridge, MA: MIT Press.

\bibitem[{Rougier(2007)}]{Rougier}
Rougier, J.~C. (2007), \enquote{Lightweight emulators for multivariate
  deterministic functions,} Tech. Rep. 07/02, MUCM Technical Report, University
  of Durham.

\bibitem[{Sacks et~al.(1989)Sacks, Welch, Mitchell, and Wynn}]{Sacks}
Sacks, J., Welch, W.~J., Mitchell, T.~J., and Wynn, H.~P. (1989),
  \enquote{Design and analysis of computer experiments (with discussion),}
  \textit{Statistical Science}, 4, 409--435.

\bibitem[{Santner et~al.(2003)Santner, Williams, and Notz}]{Santner}
Santner, T.~J., Williams, B.~J., and Notz, W.~I. (2003), \textit{The Design and
  Analysis of Computer Experiments}, New York: Springer.

\bibitem[{Tang(1993)}]{Tang}
Tang, B. (1993), \enquote{Orthogonal array-based {L}atin hypercubes,}
  \textit{Journal of the American Statistical Association}, 88, 1392--1397.

\bibitem[{Taylor and Lane(2004)}]{TaylorLane}
Taylor, B. and Lane, A. (2004), \enquote{Development of a novel family of
  military campaign simulation models,} \textit{Journal of the Operational
  Research Society}, 55, 333--339.

\end{thebibliography}

\end{document}